\documentclass[aps,prl,twocolumn,superscriptaddress,showpacs]{revtex4}
\usepackage{graphicx}
\usepackage{dcolumn}

\begin{document}

\title{Structure of $^{240}$Pu: Evidence for Octupole Phonon Condensation?}

\author{X. Wang}
\altaffiliation{Present address: Department of Physics, Florida State University, Tallahassee, FL 32306}
\affiliation{Physics Division, Argonne National Laboratory, Argonne, IL 60439}
\affiliation{Physics Department, University of Notre Dame, Notre Dame, IN 46556}
\author{R. V. F. Janssens}
\author{M. P. Carpenter}
\author{S. Zhu}
\affiliation{Physics Division, Argonne National Laboratory, Argonne, IL 60439}
\author{I. Wiedenh{\"o}ver}
\affiliation{Department of Physics, Florida State University, Tallahassee, FL 32306}
\author{U. Garg}
\author{S. Frauendorf}
\affiliation{Physics Department, University of Notre Dame, Notre Dame, IN 46556}
\author{T. Nakatsukasa}
\affiliation{RIKEN Nishina Center, 2-1 Hirosawa, Wako, Saitama 351-0198, Japan}
\author{I. Ahmad}
\affiliation{Physics Division, Argonne National Laboratory, Argonne, IL 60439}
\author{A. Bernstein}
\author{E. Diffenderfer}
\affiliation{Department of Physics, Florida State University, Tallahassee, FL 32306}
\author{S. J. Freeman}
\affiliation{Physics Division, Argonne National Laboratory, Argonne, IL 60439}
\affiliation{Schuster Laboratory, University of Manchester, Manchester M13 9PL, UK}
\author{J. P. Greene}
\author{T. L. Khoo}
\affiliation{Physics Division, Argonne National Laboratory, Argonne, IL 60439}
\author{F. G.  Kondev}
\affiliation{Nuclear Engineering Division, Argonne National Laboratory, Argonne, IL 60439}
\author{A. Larabee}
\affiliation{Greenville College, Greenville, IL 62246}
\author{T. Lauritsen}
\author{C. J. Lister}
\affiliation{Physics Division, Argonne National Laboratory, Argonne, IL 60439}
\author{B. Meredith}
\affiliation{Greenville College, Greenville, IL 62246}
\author{D. Seweryniak}
\affiliation{Physics Division, Argonne National Laboratory, Argonne, IL 60439}
\author{C. Teal}
\author{P. Wilson}
\affiliation{Department of Physics, Florida State University, Tallahassee, FL 32306}

\date{\today}

\begin{abstract}

The expanded level structure of $^{240}$Pu available from
the present study highlights the role of strong octupole correlations
in this nucleus.  Besides a delayed alignment in the yrast band, the
observations include the presence of both
$I^{+}{\rightarrow}(I-1)^{-}$ and $I^{-}{\rightarrow}(I-1)^{+}$ E1
transitions linking states of the yrast and negative-parity bands at
high spin and the presence of an additional even-spin, positive-parity band
deexciting exclusively to the negative parity sequence.  The
observations appear to be consistent with expectations based on the
recently proposed concept of octupole phonon condensation.

\end{abstract}


\maketitle

Octupole correlations in nuclei result from the long-range, octupole-octupole interaction between nucleons occupying
pairs of orbitals with $\Delta{j}=\Delta{l}=3$.  When both valence protons and neutrons occupy such states, the
strength of the correlations can be such that rotational bands with alternating parity appear.  These have commonly
been interpreted in terms of the rotation of octupole-deformed nuclei~\cite{Ahmad-ARNPS-43-71-93,Butler-RMP-68-349-96}.
The most striking examples of such behavior have been found in lanthanide ($A\sim 146$) and actinide ($A\sim 224$)
nuclei.  However, it has been realized for some time that this picture does not account for all the experimental
observations.  Typically, the positive and negative parity states merge only at high spin.  Moreover, a single
rotational sequence rarely develops.  Rather, the negative- and positive-parity sequences cross on close-lying
trajectories in the energy-spin plane.  An alternative interpretation for this behavior has recently been proposed in
Ref.~\cite{Frauendorf-PRC-77-021304R-08} where these states are understood as resulting from rotation-induced
condensation of octupole phonons having their angular momentum aligned with the rotational axis.  When the rotation of
the condensate and the quadrupole shape of the nucleus synchronize, the collective motion becomes the familiar rotation
of a static octupole shape. The small deviations mentioned above indicate that the synchronization is not fully
reached. For less-well deformed nuclei, the resulting collective motion resembles that of a reflection-asymmetric tidal
wave traveling on the surface of the nucleus~\cite{Reviol-PRC-74-044305-06}.

In contrast to the situation depicted above, most negative-parity bands in the $A\sim 230-250$ actinide nuclei are
described satisfactorily as rotational bands built on octupole vibrations.  In this context, the $^{238-240}$Pu
isotopes have remained somewhat of a puzzle.  Specifically, Wiedenh{\"o}ver
$\it{et~al.}$~\cite{Wiedenhover-PRL-83-2143-99} reported that the sharp $i_{13/2}$ proton alignment observed in the
yrast sequence of $^{241-244}$Pu is absent within the same frequency range in the three lighter isotopes. This
observation implies that, at the very least,  there is a significant delay in the alignment process (octupole
deformation has been shown to delay alignment processes - see Ref.~\cite{Frauendorf-PLB-141-23-84}). Furthermore, at
the highest spins, the yrast and the octupole states in $^{238,240}$Pu appear to merge into a single band, although the
interleaving E1 dipole transitions between the two sequences could not be observed.  In $^{239}$Pu, the positive- and
negative-parity bands merge at high spin as well, and parity-doublets seem to appear~\cite{Zhu-PLB-618-51-05}.
Additional indications for strong octupole correlations come from the fact that the decoupling parameters for the
$K^{\pi}=1/2^+$ and $K^{\pi}=1/2^-$ bands in $^{239}$Pu are of similar magnitude, but opposite in
sign~\cite{Zhu-PLB-618-51-05} and, in the three isotopes, large dipole moments were inferred from the measured
$B(E1)/B(E2)$ ratios in the medium-spin range where they are
available~\cite{Wiedenhover-PRL-83-2143-99,Zhu-PLB-618-51-05}. Finally, it is worth noting that the large octupole
strength in these specific Pu nuclei also manifests itself in the properties of their $\alpha$
decay~\cite{Sheline-PRC-61-057301-00}.

This letter focuses on $^{240}$Pu, the even Pu isotope with the strongest octupole correlations.  A new, high
statistics measurement has expanded considerably the level structure of this nucleus, and uncovered additional,
unexpected spectral consequences of these correlations.  All the observations appear to find a satisfactory
interpretation within the framework of octupole phonon condensation.

\begin{figure}
\begin{center}
\includegraphics[angle=0,width=0.70\columnwidth]{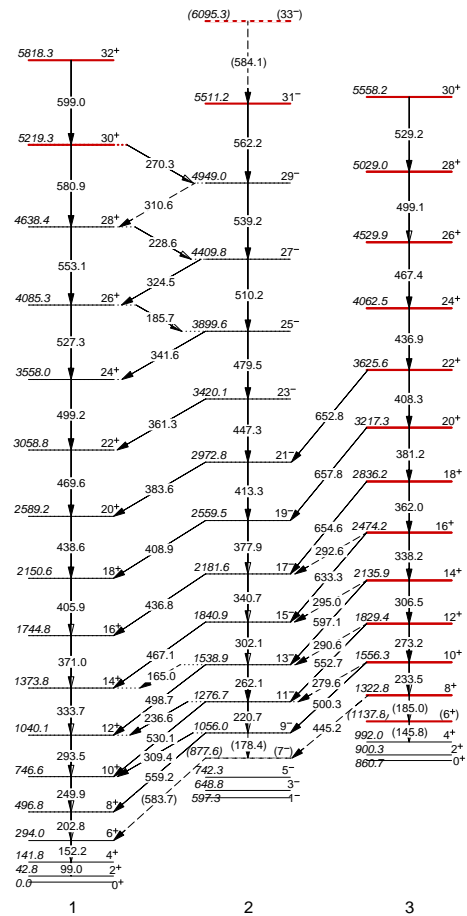}
\caption {(Color online) Level scheme of $^{240}$Pu with the new states observed in this work indicated by thick red
lines; dashed lines and transitions energies under parentheses are used for $\gamma$ rays with either a tentative
placement, or with no angular correlation information.
\label{fig:Pu240_levl_sche}}
\end{center}
\end{figure}

The experiment was carried out at the ATLAS accelerator facility at Argonne National Laboratory, with the Gammasphere
array of 101 Compton-suppressed Ge spectrometers~\cite{Lee-NPA-520-641c-90}.  The approach is identical to that
described in Ref.~~\cite{Wiedenhover-PRL-83-2143-99}.  Thus, the so-called ``unsafe'' Coulomb excitation
technique~\cite{Ward-NPA-600-88-96} was used at beam energies above the Coulomb barrier to enhance
the feeding of the highest-spin states and the population of weak $\gamma$-ray cascades.  A $^{208}$Pb beam at an
energy of 1300 MeV from ATLAS bombarded a $^{240}$Pu target of $\sim$0.35 mg/cm$^2$ thickness electroplated onto a
50-mg/cm$^2$ Au backing~\cite{Greene-NIMA-438-119-99}.  About 3$\times$10$^9$ coincidence events were recorded when
three or more suppressed Ge detectors fired in prompt coincidence.  In the subsequent analysis, the data were sorted
into three- and four-dimensional histograms using the Radware~\cite{Radford-NIMA-361-297-95} and Blue
database~\cite{Cromaz-NIMA-462-519-01} analysis packages.  The latter made important contributions because of the
ability to produce coincidence spectra at different detector angles which proved essential for the angular correlation
measurements, as well as for the delineation of the highest-spin states where Doppler shifts and/or broadenings
complicate the analysis.  The resulting $^{240}$Pu level scheme of Fig.~\ref{fig:Pu240_levl_sche} is based on the
observed coincidence relationships, as well as the measured angular correlations and $\gamma$-ray intensities.  Further
details of the data analysis and of the results will appear in a forthcoming publication which will also include data
on $^{238,242}$Pu~\cite{Wang-Pus-unpublsh-07}. They can also be found in Ref.~\cite{Wang-07-thesis}.

With respect to
Refs.~\cite{Hseuh-PRC-23-1217-81,Parekh-PRC-26-2178-82,Hardt-NPA-407-127-83,Hackman-PRC-57-R1056-98,Wiedenhover-PRL-83-2143-99},
the highest-spin levels in the yrast and first negative-parity bands (Bands 1 and 2 in Fig.~\ref{fig:Pu240_levl_sche})
in $^{240}$Pu have been delineated more accurately, hereby confirming the absence of any significant departure from a
straightforward rotational behavior up to spins in excess of $30\hbar$ reported
earlier~\cite{Wiedenhover-PRL-83-2143-99}.  Because of the high statistics accumulated, all the levels of odd spin and
negative parity up to 29$^-$ have now been linked to the positive-parity yrast states by $E1$ transitions of the
$I^{-}{\rightarrow}(I-1)^{+}$ type.  More importantly, above spin 24$\hbar$, where Bands 1 and 2 appear to merge in a
single sequence, three linking transitions of the type $I^{+}{\rightarrow}(I-1)^{-}$, $\it{i.e.}$, the 185.7-, 228.6-
and 270.3-keV dipole transitions have also been observed, establishing for the first time in a Pu isotope the
interleaving pattern of $E1$ transitions commonly associated with octupole rotation. Further evidence for strong
octupole correlations comes from the branching ratios between out-of-band and in-band transitions in Band 2. For
$I$$\leq$23$\hbar$, these ratios were presented in Fig. 3 of Ref.~\cite{Wiedenhover-PRL-83-2143-99}. They have now been
extended throughout Band 2, and their rise with angular momentum was found to persist to the
highest spins (see Fig. 4.39 in Ref.~\cite{Wang-07-thesis}). With the usual assumption of a constant quadrupole moment
$Q_0=11.6\;eb$ adopted from the measured $B(E2; 2^{+}{\rightarrow}0^{+})$ transition rate~\cite{Bemis-PRC-8-1466-73},
the induced intrinsic dipole moments $D_0$ are large: $D_0{\geq}0.2\;efm$ for $I{\geq}25\hbar$, and correspond to rates
$B(E1){\geq}2{\times}10^{-3}$ W.u. Such values are larger than the $B(E1)$ strengths commonly observed for transitions
linking negative-parity levels to positive-parity yrast states ($<10^{-4}$ W.u.). They are, however, of the same order
as those reported in the light Ra and Th isotopes~\cite{Schuler-PLB-174-241-86}, nuclei often viewed as some of the
best examples of octupole rotors~\cite{Butler-RMP-68-349-96}.

Built on three levels established previously in the decay of $^{240}$Np~\cite{Parekh-PRC-26-2178-82}, Band 3 of
Fig.~\ref{fig:Pu240_levl_sche} has been extended up to $I^{\pi}$ = 30$^+$. Most surprisingly, from the 8$^+$ state
upward, the deexcitation of this band proceeds solely towards the negative-parity levels of Band 2 with
$I^{+}{\rightarrow}(I-1)^{-}$ dipole transitions of $E1$ character as well as with weaker $I^{+}{\rightarrow}(I+1)^{-}$
$\gamma$ rays~\cite{Wang-Pus-unpublsh-07,Wang-07-thesis}. The $E1$ multipolarity was established from the angular
correlation analysis~\cite{Wang-Pus-unpublsh-07,Wang-07-thesis} showing that the linking transitions exhibit negative
$A_2$ coefficients and small, positive $A_4$ coefficients (consistent with zero) of the same magnitude than those
measured for the $E1$ transitions linking Bands 2 and 1~\footnote{For example, the 552.7 keV $12^+\rightarrow 11^-$
transition is characterized by the correlation coefficients $A_2/A_0=-0.23(4)$ and $A_4/A_0=0.03(3)$.}. No linking
transitions between Bands 1 and 3 were observed. For example, while the 14$^+$ Band 3 member decays to the 13$^-$ level
of Band 2 by a transition with an intensity of $\sim$0.7$\%$ of the 249.9-keV yrast transition, the corresponding upper
limits for linking transitions to either the 12$^+$ or 14$^+$ Band 1 states is $\leq$0.1$\%$. This is a surprising and,
to the best of our knowledge, unique observation in deformed nuclei throughout the periodic table. The $E1$ transitions
linking Bands 3 and 2 are similar in character to those linking Bands 2 and 1: within the same spin range the
$B(E1)/B(E2)$ branching ratios between out-of-band and in-band transitions for the former are of the same magnitude
within errors than those observed for the latter (and the associated $D_0$ moments are large). Note that this result
provides an explanation for the findings of Hoogduin {\it et al.}~\cite{Hoogduin-PLB-384-43-96} who reported that,
while in neighboring even nuclei strong $E0$ transitions link the first excited even-spin, positive-parity band to the
yrast states, none was found for the band based on the 861-keV $0^+_2$ level of $^{240}$Pu. Rather, in this case strong
$E0$ deexcitation is associated with a rotational sequence built on the 1091-keV $0^+_3$ state not observed in the
present measurements.

\begin{figure}
\begin{center}
\includegraphics[angle=-90,width=0.58\columnwidth]{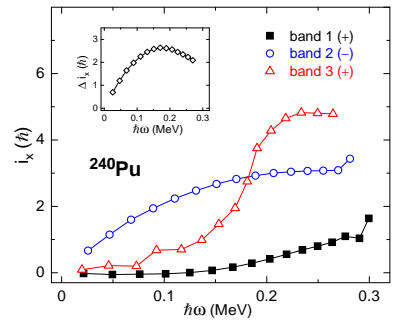}
\caption{(Color online) Aligned spins, $i_x$, as a function of rotational frequency deduced for Bands 1, 2 and 3 in
$^{240}$Pu. A collective reference with the Harris parameters ${\cal J}_0=71\hbar^2$MeV$^{-1}$ and ${\cal
J}_1=320\hbar^4$MeV$^{-3}$ is adopted, which ensures that Band 2 becomes flat for $\hbar \omega > 0.2$ MeV.
The inset provides the relative alignment $\Delta{i}_{x}$ of Band 2 with respect to the yrast sequence.
\label{fig:240Pu_alignment_delX}}
\end{center}
\end{figure}

\begin{figure}
\begin{center}
\includegraphics[angle=-90,width=0.58\columnwidth]{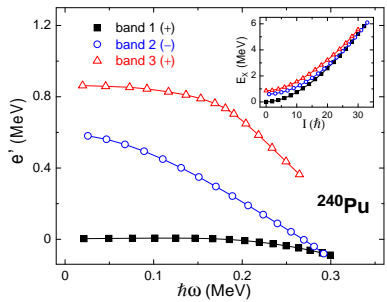}
\caption{(Color online) Routhians ($e^{\prime}$) as a function of angular frequency ($\hbar\omega$) for states in Bands 1, 2 and 3 of
$^{240}$Pu. The same reference as in Fig. \ref{fig:240Pu_alignment_delX} is used.
Insert: The excitation energies ($E_x$) as a function of spin ($I$) for the
same bands. \label{fig:240Pu_ex_ener_routhian}}
\end{center}
\end{figure}

In order to gain further insight into the nature of the excitations in $^{240}$Pu, the evolution with rotational
frequency $\hbar\omega$ of their alignments $i_x$ and routhians $e'$ are displayed in
Figs.~\ref{fig:240Pu_alignment_delX} and~\ref{fig:240Pu_ex_ener_routhian}, respectively. The
striking absence of the strong $i_{13/2}$ proton alignment at $\hbar$$\omega$$\sim$ 0.25 MeV seen in
$^{242,244}$Pu~\cite{Wiedenhover-PRL-83-2143-99} is confirmed in the yrast sequence (Band 1). The highest-frequency
data may suggest that a first alignment is, in fact, delayed as expected in a nucleus with octupole
deformation~\cite{Frauendorf-PLB-141-23-84}, but an extension of Band 1 to higher spins is required before a firm
conclusion can be reached. The alignment of Band 2 grows gradually from a small value (${\sim}0.5\hbar$) at
$\hbar\omega {\sim}$ 0.02 MeV to a maximum of $\sim$3$\hbar$ at $\hbar\omega{\sim}$ 0.20 MeV, before remaining essentially
constant up to $\hbar\omega {\sim}$ 0.28 MeV. As a result, the value of $\Delta{i}_{x}$, the relative alignment of Band 2
with respect to the yrast sequence (insert in Fig.~\ref{fig:240Pu_alignment_delX}), reaches 3$\hbar$ in the medium spin
region, as expected for the alignment of an octupole phonon. Furthermore, a small decrease in $\Delta{i}_{x}$ occurs
for $\hbar\omega>$ 0.20 MeV, a phenomenon also observed in nuclei such as $^{220}$Ra and
$^{222}$Th~\cite{Smith-PRL-75-1050-95,Cocks-NPA-645-61-99} that are often considered to be among the best
``octupole-deformed'' rotors. Thus, the similarities between these rotors and Bands 1 and 2 at high spin extend beyond
the interleaving of the states in the two sequences accompanied by the characteristic pattern of $E1$ interband
transitions to the more subtle behavior of relative alignments.

The aligned spin of Band 3 (Fig.~\ref{fig:240Pu_alignment_delX}) starts from a low value ($<1\hbar$) and a first, small
irregularity in the evolution of $i_x$ with $\hbar\omega$ occurs before 0.1 MeV, signaling the presence of a band
crossing that can also be traced in the routhian of Fig.~\ref{fig:240Pu_ex_ener_routhian}. More importantly, at the
point where a $3\hbar$ alignment is achieved in Band 2, $\it{i.e.}$, $\hbar\omega {\sim}$ 0.20 MeV, a gain of
${\sim}5\hbar$ has occurred in Band 3, an $i_x$ value that remains essentially constant at higher frequencies. This
alignment gain translates into a slope change for the routhian of Band 3, and the ensuing trajectory would likely
result in a crossing with the routhians of Bands 1 and 2 at frequencies beyond those reached in this work.

The recently proposed scenario of condensation of rotational-aligned octupole phonons
~\cite{Frauendorf-PRC-77-021304R-08} explains the observations in a natural way. Briefly, in this framework a prolate
nucleus rotates with a frequency $\omega_2$, while multi-phonon octupole tidal waves travel over its surface with a
frequency $\omega_3$, which is 1/3 of the vibrational frequency (see Fig. 4 in
Ref.~\cite{Frauendorf-PRC-77-021304R-08}). With increasing angular momentum, the one-phonon ($n=1$) band crosses the
zero-phonon ($n=0$) band and becomes yrast. Then the two-phonon ($n=2$) band becomes lowest in excitation energy
following crossings with the $n=0$ and $n=1$ sequences. In the same way, the $n=3$ phonon band comes down toward the
yrast line as the angular momentum increases, $\it{i.e.}$, boson condensation occurs; see Fig. 3(b) in
Ref.~\cite{Frauendorf-PRC-77-021304R-08}. In the ideal case of non-interacting harmonic phonons, these crossings occur
at the same critical frequency $\omega_{c}=\omega_2=\omega_3$, where the routhians of multi-phonon states coincide. At
$\omega_{c}$, the condensate co-rotating with the prolate nucleus resembles in a striking way an octupole rotor. In
contrast to a static octupole rotor, $\omega_2$ and $\omega_3$ are not locked, and the lowest positive- and
negative-parity bands oscillate around the ideal condensation line (Fig. 3(c) in
Ref.~\cite{Frauendorf-PRC-77-021304R-08}). The anharmonicities of the octupole mode lead to an interaction between
states with different phonon numbers but the same parity, resulting in a repulsion and mixing of the crossing bands.

Within this framework, Bands 1, 2 and 3 in $^{240}$Pu are associated at the highest rotational frequencies with $n=0$,
$n=1$ and $n=2$ phonons and respective $i_x$ values of approximately 0, 3 and $6\hbar$ result from the alignment of the
octupole phonons with the rotation axis. The critical frequency is $\hbar \omega_c\approx 0.3$ MeV, where the $n=0$ and
$n=1$ bands (Bands 1,2) cross in Fig. \ref{fig:240Pu_ex_ener_routhian}. The routhians of the $n=0$ and
$n=2$ bands (Bands 1, 3) do not quite come together because they are associated with bands of the same parity. Due to
the interaction, they repel each other. The corresponding state mixing gradually increases the $i_x$ value of the $n=0$
band by about $1\hbar$  and reduces  $i_x$  of the $n=2$ band to about 5$\hbar$.  The similarity between the
experimental routhians $e^{\prime}$ and excitation energies $E_x$ of the three bands
(insert in Fig.~\ref{fig:240Pu_ex_ener_routhian}) and the expectations presented in  Figs. 3(b) and 3(c) of
Ref.~\cite{Frauendorf-PRC-77-021304R-08} is notable. Although the routhian of the $n=2$ band (Band 3) could not be
extended to the frequencies required to delineate the crossing and interaction with Band 1, the data seem to indicate
that such a crossing would likely occur above 0.3 MeV, pointing to the presence of anharmonicities. The interaction
of the low-lying $n=2$ band with the $n=0$ band explains why the proton $i_{13/2}$ alignment is delayed (or absent). It
is responsible for the down bend in the trajectory of the routhian of Band 1 at the highest frequencies, and will delay
a crossing with the s-band (band with two aligned $i_{13/2}$ protons). More importantly, the $n=2$ band will
strongly interact with this s-band and push it
up in excitation energy, delaying its crossing with Band 1.

The interpretation of Band 3 as a two-phonon octupole structure above $\hbar\omega\sim$ 0.1 MeV provides a natural
explanation for the sole presence of $E1$ transitions between Bands 3 and 2 as only transitions between states
differing by a single phonon are allowed within such a vibrational picture. The mixing with the zero-phonon band at
high spin accounts for the $E1$ transitions linking Bands 1 and 2, with the presence of both
$I^{-}{\rightarrow}(I-1)^{+}$ and $I^{+}{\rightarrow}(I-1)^{-}$ $\gamma$ rays of comparable strength
~\cite{Frauendorf-PRC-77-021304R-08,Wang-07-thesis}. As pointed out in
Refs.~\cite{Frauendorf-PRC-77-021304R-08,Wang-07-thesis}, a substantial mixing of zero- and two-phonon bands does not
differ much from a static octupole deformation.

In summary, the new data available for $^{240}$Pu illustrate the impact of strong octupole correlations in ways that
had not been seen before. A satisfactory interpretation of the observations appears to be achieved with the recently
proposed concept of octupole condensation. To validate this interpretation further, additional experimental and
theoretical work is highly desirable. First, extending studies of $^{240}$Pu to higher spins would provide missing
information about the predicted crossing and interaction of Bands 1 and 3. More detailed information on the
electromagnetic transition rates will be important to substantiate this interpretation further. A search for the same
phenomena in other nuclei, not only in the actinide region, but also through other regions of the nuclear chart is
important in order to assess the generality of this exotic mode. Finally, a full microscopic description  of this
collective behavior is required as well.

This work is supported in part by the U.S. Department of Energy, Office of Nuclear Physics, under contract No.
DE-AC02-06CH11257 and grant DE-FG02-95ER40934 and the U.S. National Science Foundation under grants No. PHY04-51120
and PHY07-54674, as well as by the ANL-Notre Dame Nuclear Theory Initiative.


\end{document}